\documentstyle[epsfig,aps,prl,twocolumn]{revtex}

\begin{document}

\title{Plasma Oscillations and Expansion of an Ultracold Neutral Plasma }

\author{S. Kulin, T. C. Killian, S. D. Bergeson\cite{scott}, and S. L. Rolston}

\address{National Institute of Standards and Technology, 
Gaithersburg, MD 20899-8424\\
(Accepted \it{Phys. Rev. Lett.})}

\maketitle

\begin{abstract}
We report the observation of plasma oscillations in an ultracold
neutral plasma. With this collective mode we  probe 
the electron density distribution and study the expansion
of the plasma as a function of time. 
For classical plasma conditions, {\it i.e.} weak 
Coulomb coupling, the expansion 
is dominated by the pressure of the 
electron gas and is  described by a hydrodynamic model.
Discrepancies between the model and observations 
at low temperature and high density may be due to strong coupling of 
the electrons.

\end{abstract}
\narrowtext

One of the most interesting features of neutral plasmas is the
rich assortment of collective modes that they support.
The most common of these is the
plasma oscillation \cite{tla29},  in which  electrons oscillate  
around their equilibrium positions
and ions are essentially stationary. 
This mode is a valuable probe of ionized gases
because 
the oscillation frequency depends solely on
the electron density.

In an ultracold neutral plasma as reported in \cite{kkb99},
the density is nonuniform and changing in time. 
A diagnostic of 
the density is thus necessary for  a  variety of experiments,
such as  determination of  the three-body recombination rate at 
ultralow temperature
\cite{hah97}, and 
observation of the effects  of strong Coulomb coupling \cite{ichimaru82}
in a two-component system.
A density probe would also aid in the study of the evolution 
of a dense gas of cold Rydberg atoms 
to a  plasma \cite{nnp99}, which may be an analog of the  
Mott insulator-conductor phase 
transition \cite{mott}.

In this work we excite 
 plasma oscillations in an ultracold 
neutral plasma by applying a radio frequency (rf) 
electric field. The  oscillations are used to map  
the  plasma density distribution and  reveal the particle dynamics
and energy flow during the expansion of the ionized gas. 

The creation of an ultracold plasma has been described  in 
\cite{kkb99}. A few million metastable xenon atoms are laser cooled 
to approximately $10\,\mu$K. The peak density  is about $2\times 
10^{10}$\,cm$^{-3}$ and the spatial distribution of the cloud 
is Gaussian with an rms radius $\sigma \approx 220\,\mu$m. 
These parameters are determined with resonant laser absorption imaging
\cite{matt}. 
To produce the plasma,
up to $25$\% 
of the atoms  are photoionized   in a two-photon excitation. Light for 
this process is provided by a Ti:sapphire laser  at $882$\,nm and  a 
 pulsed dye laser  at $514$\,nm ($ 10\,$ns pulse length). 
Because of the small electron-ion mass ratio, the resulting
electrons have an initial kinetic energy ($E_{e}$) approximately equal to the 
difference between the photon energy and the  ionization potential. 
In this study we vary $E_{e}/k_{B}$ between 1 and 1000\,K.
The initial kinetic energy of the ions varies between 10\,$\mu$K and 
4\,mK.

For detection of charged particles,
a small  DC field (about $1$\,mV/cm) directs
electrons to a single channel electron multiplier and ions 
to a multichannel plate detector. 
The amplitude of 
the rf field that excites  plasma oscillations,
$F$, varies between $0.2-20\,$mV/cm rms.
All electric fields are applied to the plasma with 
grids located
above and below the laser-atom interaction region.

In the absence of a magnetic field,
the frequency of plasma oscillations is given by
$f_{e}=(1/2\pi)\sqrt{e^{2}n_{e}/\epsilon_{0}m_{e}}$ \cite{tla29}. 
Here,  $e$ is the elementary 
charge,  $n_{e}$ is the electron density,
$\epsilon_{0}$ is the permittivity of vacuum, and 
$m_{e}$ is the  electron mass. 
This  relation is most often derived for an infinite 
homogeneous plasma, but it is also valid in our inhomogeneous system 
for modes which
are localized in regions of near resonant density. 
Corrections to $f_{e}$ due to finite temperature \cite{bgr49} 
depend on the wavelength of 
the collective oscillation, 
which is difficult to accurately estimate.
Such corrections are not expected to be large and 
will be  neglected.
We observe plasma oscillations with frequencies from $1$ to 
$250$\,MHz. This corresponds to 
resonant electron densities, $n_{r}$,
between $1\times10^{4}$\,cm$^{-3}$ and
$8 \times10^{8}$\,cm$^{-3}$. 
The oscillation frequency is  sensitive only to $n_{e}$,
but, as explained in \cite{kkb99}, the core of the plasma
is neutral. This  implies that 
plasma oscillations
measure   electron and ion densities in this region ($n_{e}=n_{i}\equiv n$).

Figure \ref{compositepaper}a shows electron signals
from an ultracold 
neutral plasma created by photoionization at time $t=0$.
Some electrons leave the sample and 
arrive at the detector at about $1\,\mu$s, producing the first
peak in the  signal. 
The resulting excess positive charge in the plasma creates 
a Coulomb potential well that traps the remaining electrons \cite{kkb99}.
In the work reported here, typically 
$90-99$\% of the electrons are trapped. 
Debye shielding maintains local neutrality 
inside a radius $r_{e}$ beyond which
the electron density  drops to zero on a length scale equal to  
$\lambda_D$.  
The value  of $r_{e}$ depends on the fraction of electrons that 
has escaped, and $\lambda_D$ is the Debye screening length,
$\lambda_D =\sqrt{{\epsilon_0 k_B T_{e}}/{e^2 n_{e}}}$, where
$T_{e}$ is the electron temperature. 
For our conditions 
$r_{e}\,$\raisebox{-.6ex}{$\stackrel{>}{\sim}$}$\,2 \sigma$, and 
$\lambda_D \ll \sigma$.
As the plasma expands, the depth of the Coulomb well 
decreases,  allowing the remaining electrons to leave the trap. 
This produces the broad  peak  at $\approx 25\,\mu$s.

In the presence of an rf  field  an additional peak appears in the 
electron  signal (Fig.\ \ref{compositepaper}a). 
We  understand the generation of  this peak as follows:
The applied rf  field 
excites plasma oscillations only where the frequency is resonant. 
Energy is thus pumped into the plasma in  the  shell 
with the appropriate electron density  ($n=n_{r}$). 
The amplitude of the collective electron motion is much less than $\sigma$,
but the acquired energy is 
collisionally redistributed among all the electrons
within $10-1000$\,ns \cite{spitzer}, raising the  electron temperature. 
This increases the evaporation rate of electrons
out of the Coulomb well, which produces the plasma oscillation
response on the electron signal.

The resonant response  at a given
time, $S(t)$, is proportional to the number of electrons in the region
where the density equals $n_{r}$.
If we make  a simple local density approximation and neglect 
decoherence of the oscillations, 
$S(t) \propto F^{2}\int d^{3}r\, n({\bf r},t)\,\delta[n({\bf r},t)-n_{r}]$.
The  width in time of the observed signal (Fig.\ \ref{compositepaper}a) reflects the 
density distribution of the sample \cite{densnote}.
At early times when the density is
higher than $n_{r}$  almost everywhere, $S(t)$ is negligibly small. 
As the cloud expands and the density decreases, the response grows
because  the fraction of 
the plasma which is in resonance increases.
The peak of the response appears approximately
when the average  density, $\bar n$, becomes resonant with the rf  field. 
$S(t)$ vanishes when the peak density is less than
$n_{r}$.

The resonant response arrives later for lower
frequency (Fig.~\ref{compositepaper}b)
as expected because $\bar n$  decreases in time. 
Assuming 
that the plasma density profile remains Gaussian during the expansion, 
$S(t)$ can be evaluated and
its amplitude  scales as $F^{2}/n_{r}$.
In Fig.~\ref{compositepaper}b the data have been normalized by this factor
and the resulting  amplitudes are similar for all conditions.

By equating $\bar n$ to $n_{r}$ when the response peak arrives, we
can plot the average plasma density as a function of time (Fig.~\ref{paperalltemp}). 
The data are well described by a self similar expansion of a Gaussian cloud,
$\bar n= N/[4\pi (\sigma_0^{2}+v_{0}^{2}t^{2})]^{3/2}$, where
$\sigma_{0}$ is the  initial  rms radius 
and  $v_{0}$ is the rms radial velocity at long times. $N$ is 
determined independently by counting the number of neutral atoms with 
and without photoionization. The extracted values of $\sigma_0$ are
equal to the size of the initial  atom cloud.
In such an expansion, the average kinetic energy per particle is
$3mv_{0}^{2}/2$.

Figure \ref{expansion} shows the dependence of $v_{0}$ on density
and initial electron energy.
We first discuss data with $E_{e}\ge 70\,$K, for which 
the expansion velocities approximately follow 
$v_{0}=\sqrt{E_{e}/\alpha m_{i}}$, where $m_{i}$ is the ion mass and
$\alpha=1.7$ is a fit parameter.
For the 
plasma to expand at this rate, the ions must acquire, on average, a 
velocity characteristic of  the electron energy.
This is much greater than the initial ion thermal velocity.
Electron-ion equipartition of energy 
would yield $v_{0}=\sqrt{E_{e}/3m_{i}}$, close to
the observed value. However, due to the large electron-ion mass 
difference,   this thermalization requires milliseconds \cite{spitzer}. 
The observed expansion, in contrast,
occurs on a time scale of tens of microseconds. 
One might expect the  expansion to be dominated by the
Coulomb energy arising from the slight charge imbalance  of the 
plasma, but this energy is  about an order of magnitude less 
than the observed
expansion energy. Also, by Gauss' law, it would only be  important in the 
expansion of the non-neutral 
outer shell of the plasma. 
The oscillation probe provides information only on the neutral core
because it relies on the presence of electrons.

A  hydrodynamic model \cite{gru95}, which describes the plasma
on length scales larger than $\lambda_D$, shows that
the expansion is driven by the pressure of the electron gas.
The pressure is exerted on the ions by outward-moving electrons that are stopped and 
accelerated inward in the trap. For the hydrodynamic calculation,
ions and electrons
are treated as fluids with local densities $n_{a}({\bf r})$ and
average velocities ${\bf u}_{a}({\bf r})=\langle{\bf v}_{a}({\bf r})\rangle$.
Here, $a$ refers to either electrons or ions, and $\langle \cdots\rangle$
denotes a local ensemble average. Particle and momentum conservation
lead to the momentum balance equations
$$
m_{a}n_{a}\left[ {\partial{\bf u}_{a} \over \partial t}
 +({\bf u}_{a}\cdot \nabla){\bf 
u}_{a}\right] =-{\bf \nabla }(n_{a}k_{B}T_{a})+ {\bf R}_{ab}.\nonumber  
$$
Here $n_{a}k_{B}T_{a}$ represents  a 
scalar pressure \cite{gru95}. 
The ion and electron equations are 
coupled by  ${\bf R}_{ab}$, which is the rate of
momentum exchange between species $a$ and $b$.
The exact form of this term is unimportant for this study, but 
${\bf R}_{ab}=-{\bf R}_{ba}$. 
Plasma hydrodynamic equations typically have electric and magnetic 
field terms, but  applied and internally generated
fields are negligible when describing the expansion.

We can make a few 
simplifying approximations that are valid before the system  has significantly expanded.
The directed motion is negligible, 
so  we  set ${\bf u}_{a}\approx 0$ everywhere. 
Because $n_{e}\approx n_{i}= n$,
$\partial{\bf u}_{e}/ \partial t\approx 
\partial{\bf u}_{i}/ \partial t$. 
Due to the small electron mass,  
the rate of increase of average electron momentum 
is negligible compared to that of the ions. 
The electron momentum balance equation then yields  
${\bf \nabla }(nk_{B}T_{e})\approx {\bf R}_{ei}$, which
describes a  balance between the pressure of the electron gas and 
collisional interactions. This
is the hydrodynamic depiction of the trapping of electrons
by the ions.

In the  ion momentum balance equation, we eliminate ${\bf 
R}_{ie}$ using the electron  equation,
and we drop the pressure  term
because the ion thermal motion is negligible.  Thus
$m_{i}n\partial{\bf u}_{i} / \partial t \approx -{\bf \nabla } 
(nk_{B}T_{e})$, which 
shows that the pressure of the electron gas  drives the 
expansion\cite{expnote}. 
This result implies that the 
ions acquire a velocity of order $\sqrt{k_{B}T_{e}/m_{i}}$, which is 
in qualitative agreement with the high $E_{e}$  data of Fig.~\ref{expansion}. 
To calculate the expansion velocity more quantitatively, one must consider
that as electrons move in the expanding trap, they perform work on the ions
and cool adiabatically.
The thermodynamics of this process  \cite{mmf93} is
beyond the scope of this study.

The data in Fig.\  \ref{expansion} indicate  that about $90$\,\% of the
initial kinetic energy of the electrons is transferred to the
ions'   kinetic energy,  $3m_{i} 
v^{2}_{0}/2=3E_{e}/2\alpha$. 
This does not imply that the {\it temperature} of the ions becomes
comparable  to $E_{e}/ k_{B}$ in this process. For the ions, ${\bf u}_{i}$ increases, 
but $m_{i}\langle|{\bf v}_{i}-{\bf u}_{i}|^{2}\rangle$,
which  measures random thermal motion
and thus temperature, is expected to
remain small.  This follows from 
 slow ion-electron thermalization \cite{spitzer} and 
correlation between position and
velocity during the expansion \cite{ows99}.

We now turn our attention to systems with $E_{e}<70\,$K (Fig.\ \ref{expansion}). They 
expand  faster than expected from  an extrapolation of
$v_{0}=\sqrt{E_{e}/\alpha m_{i}}$, and thus do not even qualitatively follow the
hydrodynamic model. 
A relative measure
of the deviation is $(m_{i}v_{0}^{2}-E_{e}/\alpha)/(E_{e}/\alpha)$.
Figure \ref{expheatstudy} shows  that the relative
deviation increases with increasing 
electron Coulomb coupling parameter \cite{ichimaru82}, 
$
\Gamma_{e} = (e^{2} / 4\pi \varepsilon_{0}\,a) / k_{B}T_{e}.   
$ 
Here,  $a=(4\pi n/3)^{-1/3}$ is the Wigner-Seitz radius,
$n$ is the  peak density at  $t=0$, and
the temperature is calculated by 
$3k_{B}T_{e}/2=E_{e}$.

The fact that the relative deviation depends only on $\Gamma_{e}$,
and that it becomes significant 
as $\Gamma_{e}$ approaches 1, 
suggests that we are observing the effects of strong 
coupling of the electrons \cite{ioncoupling}. The hydrodynamic model of the plasma 
is only valid when $\Gamma _{e} \ll 1$. 
When $\Gamma _{e}\, $\raisebox{-.6ex}{$\stackrel{>}{\sim}$}$\, 1$,
electron and ion spatial distributions show 
short range correlated fluctuations that  are not accounted
for in a smooth fluid description \cite{ich92}.
Correlations between the ion and electron positions would 
provide the excess kinetic energy observed in the expansion by
lowering the potential energy of the plasma.
This satisfies overall energy conservation and it may
also explain the systematically poor fits of  the data 
for high $\Gamma_{e}$ (See Fig.\,\ref{paperalltemp}).

Strong coupling is also predicted to alter 
the relation for the frequency $f_{e}$ \cite{kgm93}, with which we extract the plasma density,
size, and expansion velocity. 
The trend of this effect  agrees qualitatively with 
the observed deviation, but knowledge of the wavelength of the 
collective oscillation  is  needed for a quantitative comparison.

Other  possible explanations for the deviation  are related to how
the ultracold plasma is created.
The $10\,$ns duration of the photoionization
pulse is long compared to the time required for 
electrons to move an interparticle spacing. Photoionization 
late in the pulse thus occurs in the presence  of free charges, which 
will depress the atomic ionization threshold by
$\Delta E_{IP}\approx \frac{1}{2} k_{B}T_{e}
\{[(3\Gamma_{e})^{3/2}+1]^{2/3}-1\}$ 
\cite{spy66}.
This effect might increase the electron kinetic energy by $\Delta 
E_{IP}$ above what has been assumed.  However, as shown in
Fig.\  \ref{expheatstudy}, the calculated $\Delta E_{IP}$
is about an order of magnitude smaller than the observed effect.
The random potential energy  of charged particles when they are 
created may also  yield a greater electron  energy than $E_{e}$ \cite{gey}.  

High $\Gamma_{e}$ (high density and low temperature)
conditions are desirable for 
studying  the three-body recombination rate in an ultracold plasma. 
The theory \cite{mke69} for this process was developed for high 
temperature, and is expected 
to break down in the ultracold regime \cite{hah97}.
Measuring  or setting an upper 
limit for the recombination rate is not possible until the dynamics of 
high $\Gamma_{e}$ systems is understood. 
We are currently studying this problem with 
molecular dynamics calculations.

We have shown that plasma oscillations are a valuable probe of the 
ion and electron density in an ultracold neutral plasma.
This tool will facilitate  future experimental studies of this novel 
system, such as
the search for other collective modes
in the plasma and    further investigation of  the effects
of correlations due to strong coupling.

We thank Lee Collins for helpful discussions and Michael Lim
for assistance with data  analysis.
S. Kulin acknowledges funding from 
the Alexander-von-Humboldt foundation. This work 
was  funded by the ONR.

\begin{center}
\epsfig{file=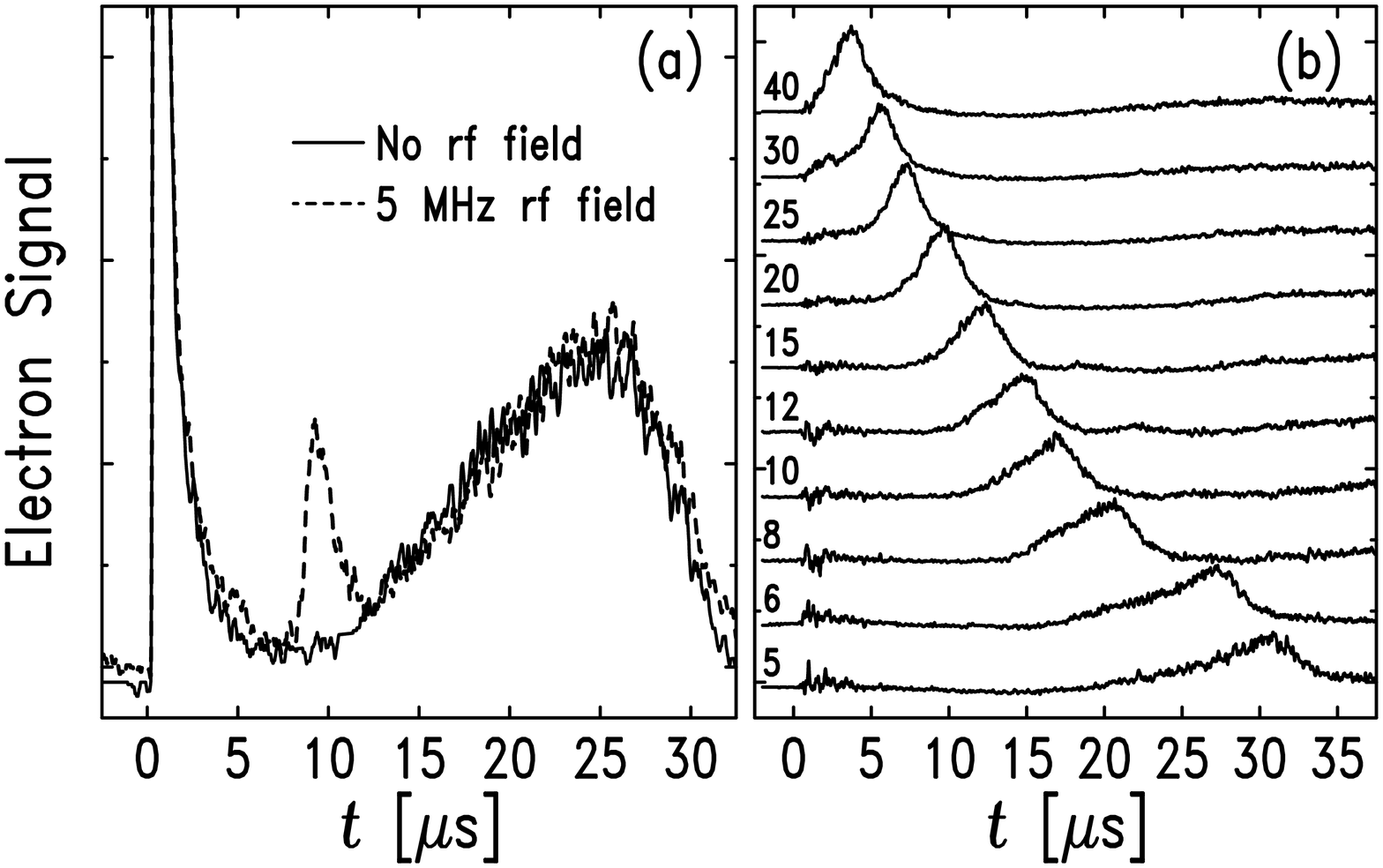, width=3.35in}
\begin{figure}[b]
\caption
{ Electron signals from  ultracold plasmas created by  photoionization
 at $t=0$. (a)  
$3 \times 10^{4}$ atoms are photoionized and
$E_{e}/k_{B}=540\,$K. Signals with and without rf field are shown.
The rf field is applied continuously.
(b) $8 \times 10^{4}$ atoms are photoionized and
$E_{e}/k_{B}=26\,$K.
For each trace, the rf frequency  in MHz is indicated, and
the nonresonant
response has been subtracted. The signals have been offset for 
clarity and have been normalized
by $F^{2}/n_{r}$. 
The resonant response arrives later for lower
frequency, reflecting expansion of the plasma.
For 40\,MHz, $n_{r}=2.0 \times 10^{7}\,$cm$^{-3}$, and
for 5\,MHz, $n_{r}=3.1 \times 10^{5}\,$cm$^{-3}$. 
}
\label{compositepaper}
\end{figure}
\end{center}

%
\begin{center}
\epsfig{file=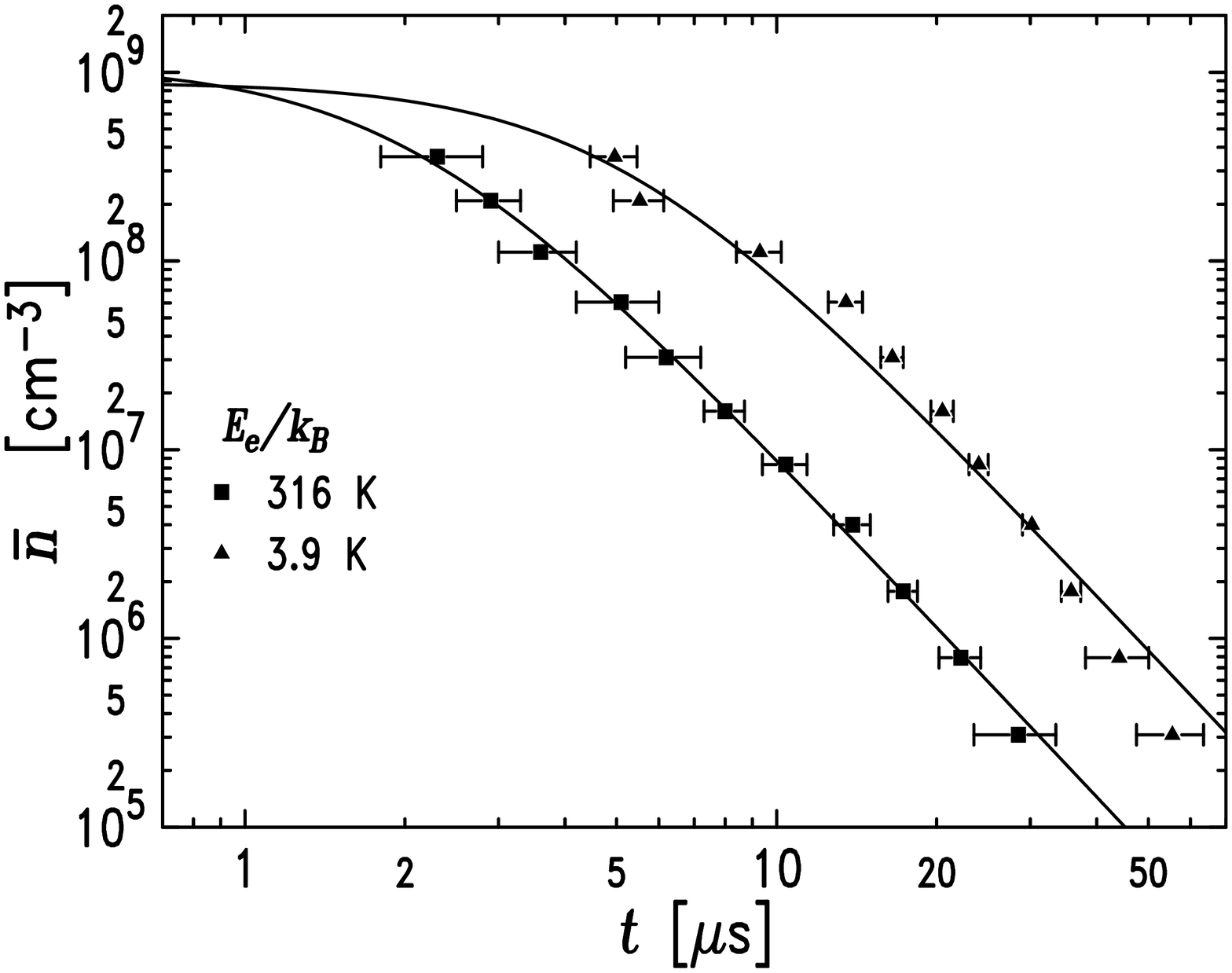, width=3.0in}
\begin{figure}
\caption
{Expansion of the plasma for $N= 5 \times 10^{5}$ photoionized atoms.
The  expansion  is  well described by 
$\bar n= N/[4\pi (\sigma_0^{2}+v_{0}^{2}t^{2})]^{3/2}$.
Horizontal  error bars arise from uncertainty in  peak 
arrival times in data
such as Fig.\ \ref{compositepaper}b.
Uncertainty in $N$ is negligible in 
this data set, but is significant for smaller $N$.
The fits 
are consistently poor at low $E_{e}$, as  in the $3.9\,$K data.
}
\label{paperalltemp}
\end{figure}
\end{center}
%
%

%
\begin{center}
\epsfig{file=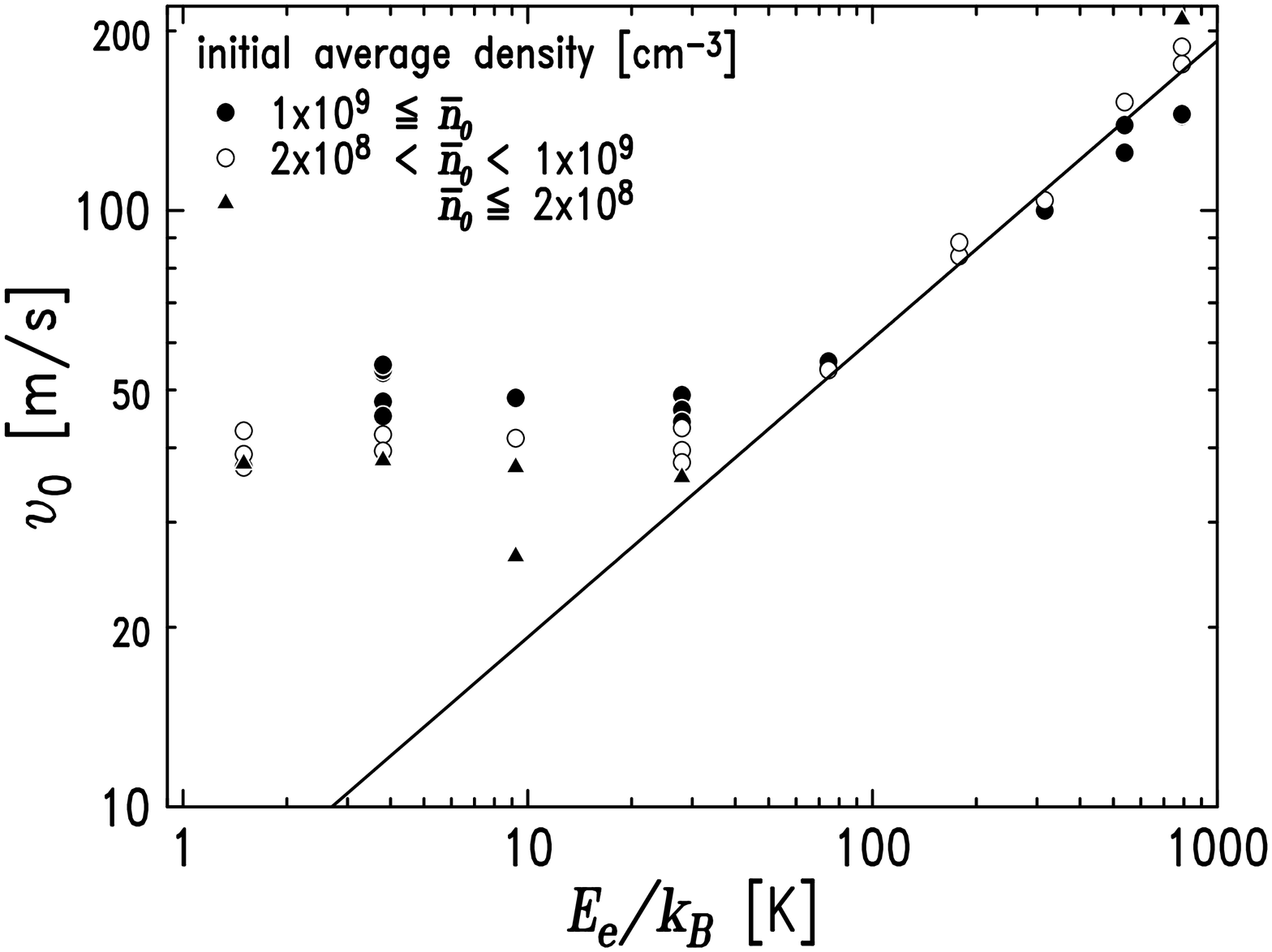, width=3.0in}
\begin{figure}
\caption
{Expansion velocities, $v_{0}$, found from fits to data such as in
Fig.\ \ref{paperalltemp}. The 
initial average density, $\bar{n}_{0}$, 
varies from $6 \times 10^{6}$ to $2.5 \times 10^{9}\,{\rm cm}^{-3}$.
The solid line, $v_{0}=\sqrt{E_{e}/\alpha m_{i}}$, with  $\alpha=1.7$,
is a fit to data with $E_{e}/k_{B}\ge 70\,$K. The behavior of low
$E_{e}$ data is discussed in the text.
Uncertainty in  $v_{0}$  
is typically equal to the size of the symbols.
There is a $0.5\,$K uncertainty in $E_{e}/k_{B}$ reflecting
uncertainty in the  dye laser wavelength. Note that for $E_{e}/k_{B}< 70\,$K,
$v_{0}$ shows a  systematic dependence on $\bar{n}_{0}$.
}
\label{expansion}
\end{figure}
\end{center}
%
%

 %
\begin{center}
\epsfig{file=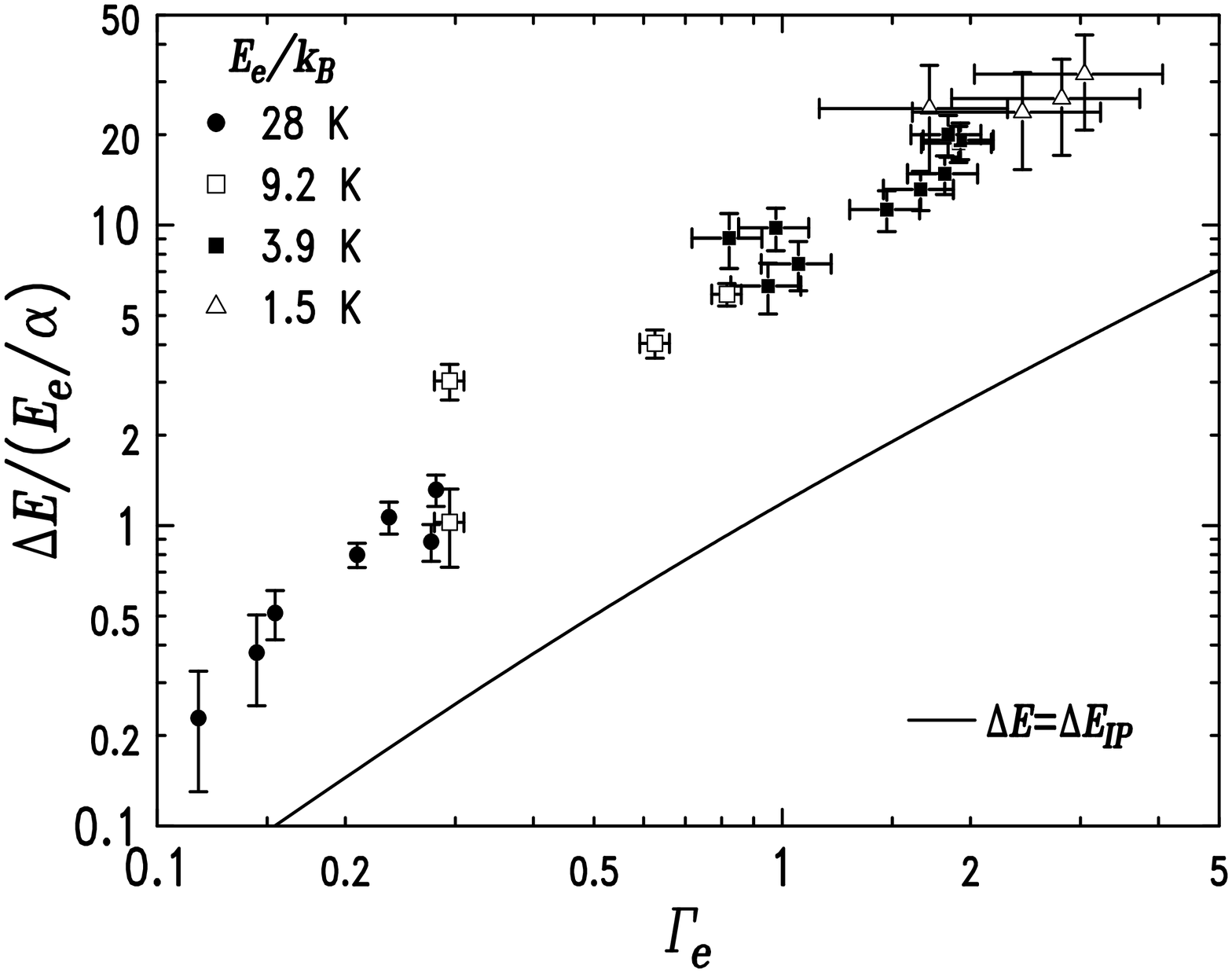, width=3.0in}
\begin{figure}
\caption
{Excess expansion energy, $\Delta E=m_{i}v_{0}^{2}-E_{e}/\alpha$, relative to  $E_{e}/\alpha$, 
as a function of $\Gamma_{e}$, the
Coulomb coupling parameter for the electrons at $t=0$. 
The solid line
results from equating $\Delta E$ to the 
predicted suppression of the atomic ionization potential in the plasma. 
Horizontal error bars arise from uncertainty in $E_{e}$. Vertical error
bars reflect uncertainty in both $E_{e}$ and $v_{0}$.}
\label{expheatstudy}
\end{figure}
\end{center}


\begin{references}
\bibitem{densnote}By applying an rf pulse,
we found the response time for excitation and detection
of  plasma oscillations to be about $1\,\mu$s. This   
is short compared to the 
width of $S(t)$ and is probably set by 
the electron thermalization rate and time-of-flight
to the detector.
\bibitem{expnote}
This equation  would also describe the ballistic expansion of a single-component 
cloud of noninteracting  particles at temperature $T_{e}$.
It also  preserves a Gaussian density distribution, 
which supports our assumption of such a distribution
in the data analysis.
\bibitem{ioncoupling} 
We expect that the ions are also strongly coupled, although we have no direct 
evidence for this. 
The  ion-ion thermalization time,  
equal to that of the electrons, is short, and the low initial ion kinetic energy
yields $\Gamma_{i}>>1$. 
Strong coupling of ions  is not expected to affect the experiments discussed here.
\end{references}
\end{document}